\documentclass[twocolumn,aps,floatfix]{revtex4}
\usepackage{amssymb}
\usepackage{graphicx}
\usepackage{amsmath}
\usepackage{pstricks}

\begin{document}
\title{Spin dynamics of two bosons in an optical lattice site: a
role of anharmonicity and anisotropy of the trapping potential}

\author{Joanna Pietraszewicz$^{1}$, Tomasz Sowi\'nski$^{1,2}$,
Miros\l{}aw Brewczyk$^{3,4}$, Maciej Lewenstein$^{2,5}$,
and  Mariusz Gajda$^{1,3}$}
\affiliation{
\mbox{$^1$Institute of Physics of the Polish Academy of Sciences,
Al.Lotnik\'{o}w 32/46, 02-668 Warszawa, Poland}\\
\mbox{$^2$ICFO - Institut de Ci\`ences Fot\`oniques, Parc Mediterrani de la Tecnologia, E-08860
Castelldefels, Barcelona, Spain} \\
\mbox{$^3$Center for Theoretical Physics of the Polish Academy of
Sciences, Al.Lotnik\'{o}w 32/46, 02-668 Warszawa, Poland}\\
\mbox{$^4$Wydzia{\l} Fizyki, Uniwersytet w Bia\l ymstoku, ul.
Lipowa 41, 15-424 Bia\l ystok, Poland} \\
\mbox{$^5$ ICREA - Instituci\'o Catalana de Recerca i Estudis Avan\c cats, 08010 Barcelona, Spain} \\
}
\date{\today}

\begin{abstract}
We study spin dynamics of two magnetic Chromium atoms trapped
in a single site of
a deep optical lattice in a resonant magnetic field.
Dipole-dipole interactions couple
spin degrees of freedom of two particles to their quantized orbital motion.
A trap geometry combined with two-body contact s-wave
interactions influence
a spin dynamics through the energy spectrum of the two atom
system. Anharmonicity and anisotropy of the site results 
in a `fine' structure of two body eigenenergies. The structure
can be easily resolved by weak magnetic dipole-dipole interactions. As an
example we examine the effect of anharmonicity and anisotropy of the
binding potential on  demagnetization processes. We show that  weak dipolar
interactions provide a perfect tool for a precision spectroscopy of the energy
spectrum of the interacting few particle system. 
\end{abstract}

\maketitle

\section{Introduction}
\label{introduction}
Dipolar magnetic interactions couple a spin of two particles to their
orbital motion in the trap \cite{Ueda_review}.
A particular example of the spin dynamics driven by  the dipole-dipole
interactions and coupled to the orbital motion of the entire
ferromagnetic sample is known as the Einstein-de Haas effect
(EdH). The effect is a macroscopic illustration of the fact that
spin contributes to the total angular
momentum of the system on the same footing as the orbital angular
momentum.

In the original experiment \cite{Richardson,EdH1,EdH2} a
rotation of a ferromagnetic rode was observed when a relative
orientation of a magnetic field (forcing a magnetization of the
sample along the field axis) and a magnetic moment of the sample
had been inverted. The 
rotation results form the fact that atomic spins adjust to a new
orientation of the magnetic field and conservation of the
$z$-component of the total angular momentum forces the orbital
angular momentum of the sample to compensate for the change of
the spin orientation. The
orbital angular momentum has to depart from its initial zero
value. The EdH effect strongly relays on the dipole-dipole
interactions. These interactions, due to their anisotropy do not
have to conserve the orbital angular momentum. Only in the
presence of the dipolar interactions initially
non-rotating system can start to rotate.

The EdH effect was also predicted for ultra cold trapped atomic
gases \cite{Ueda_1,Santos,KG}. Here it serves as a spectacular
manifestation of the dipole-dipole interactions of atomic magnetic
moments. This leads to a nontrivial dynamics coupling orbital
motion and atomic spin. Spin
dynamics due to the dipole-dipole interactions is inevitably
associated with an excitation of orbital motion of an atom in the
trap. The motion is quantized, there is therefore an energy gap
separating initial and final states of a different
magnetization. The quantization of orbital energy of atoms
significantly effects a spin dynamics. This has been noticed in
\cite{Swislocki}.

Two scenarios are possible depending on a ratio of the energy
gap to the dipolar energy. For a relatively dense gas of
polarized Chromium atoms ($10^{15} {\rm cm}^{-3}$) initially in
$S=3,m_S=3$ state and for relatively weak harmonic traps ($\omega \simeq
2\pi \times 100 {\rm Hz}$), the dipolar
energy can exceed the excitation gap what makes the spin
dynamics possible. The atoms are transfered first to $m_S=2$
state and then, due to  higher order processes
and the contact
interactions, to other magnetic components \cite{Ueda_1}.

Situation is different in optical lattices or in a case of species with a small magnetic moment,
such as ferromagnetic Rubidium.  
In general if the dipolar energy is much less than the
orbital excitation energy then spin dynamics is suppressed
due to the energy conservation.

This obstacle can be overcome. As shown in \cite{Swislocki}
the energies of both involved states can be tuned due to a
Zeeman effect. The atoms prepared in $m_S=3$ state can be
transferred to initially unoccupied $m_S=2$ components on the
expense of the Zeeman, instead of the dipolar
energy. In axially symmetric traps the phase of the $m_S=2$
atoms' wavefunction displays a characteristics vortex-like
defect signifying an orbital angular momentum in a final state.
This scenario has been studied in the mean field limit, for a
three component Rubidium spinor condensate
\cite{Swislocki}. It has been shown that different final states,
all involving vortex-like excitation, can be populated via the
dipolar processes of the EdH type. Even highly excited trap
states can be reached if energy and angular momentum
conservation are met by an appropriate choice of
the resonant magnetic field. 

Energy conservation makes the weak dipolar interactions highly sensitive, selective and
tunable by means of the external magnetic field. This requires
however, a very high precision in control of both a value and
a direction of the field applied.
Stabilization of the ultra weak magnetic field with precision below
1 mG is required for Chromium atoms \cite{Bruno_1}. The proposal
exists suggesting that the time dependent magnetic fields might
relax these constrains \cite{Brewczyk}. The resonant character of demagnetization
of initially polarized Chromium atoms in an optical lattice  
has been demonstrated recently \cite{Tolra_rezonanse}.

The EdH effect was also suggested as a mechanism to create
excited orbital states in the optical lattice
\cite{Pietraszewicz}. The mechanism leads also to an interesting
physics described by Bose-Hubbard models with two spin species
occupying different Wannier states. The axial
symmetry of a single lattice site was assumed. This assumption
seems to be quite obvious and is frequently used even in more
`exotic' situations \cite{Vincent}.
     
In this paper we want to reexamine carefully the role of the
harmonic approximation and show its consequences on the spin
dynamics, in particular for situations when higher orbital
Wannier states are involved. We study elementary processes
driven by the dipolar interactions with two atoms
trapped in a single site of a high square optical lattice.
Such a scenario was considered in \cite{You} where it has been
shown that  $100\%$ efficient resonant transfer of atoms to the
orbital state is possible. 
We show that the anharmonicity of the
binding potential plays a very important
role. In the context of ultracold atoms  the
effect of anharmonicity was discussed previously in \cite{Colin, Larson}.

The paper is organized as follows. In section \ref{model} we
define the Hamiltonian of the system as well as all
approximations and model parameters. Then in section \ref{m1} we
discuss a resonant dipolar coupling allowing for efficient spin
transfer in the channel where
spin projection of only one atom changes by one quantum. In section
\ref{m2} we present a detailed analysis of processes when spins
of both interacting atoms change by one quantum each 
(the spin of the doublon changes by two quanta). In particular we
take into account separately the effect of anharmonicity of the
trapping potential and, finally,
both effects of anisotropy and anharmonicity of a single site of
a square optical lattice. In section \ref{experiment}
we compare our results with experimental data of \cite{Tolra_rezonanse}.
Final remarks are presented in section
\ref{conclusions}.

\section{The model}
\label{model}

\subsection{The Hamiltonian}
We study a simple system of two Chromium ${}^{52}$Cr atoms in
the $S=3$ ground state in a single site of a 2D square optical
lattice created by a two pair of counter propagating laser beams
of wavelength $\lambda$ and momentum $k=2\pi/\lambda$. A
confinement in the z-direction is
provided by a harmonic potential of frequency $\Omega_z$. We
assume that the lattice depth is so big that we can neglect
tunneling to the neighboring sites. Through the paper we use the
lattice units, i.e. the distance is measured in
$1/k=\lambda/2\pi$, all energies in units of the
recoil energy $E_R = (\hbar k)^2/(2M)$, where $M$ is the atomic
mass. The units of time and frequency are $\hbar/E_R$ and
$E_R/\hbar$ respectively. The magnetic field unit is
$B_0=E_R/(g\mu_B)$, where $\mu_B$ is the Bohr magneton and $g=2$
is a Lande factor.

The Hamiltonian of the system has the following form:
\begin{align} 
\label{Hamiltonian}
{\cal H} &= \sum_{i=1}^2 \left[-\nabla_i^2
+V_0(\sin^2x_i+\sin^2y_i) +\kappa^2 z_i^2)\right] \nonumber \\
&+ \delta(\boldsymbol{r}_1-\boldsymbol{r}_2)\sum_{{\cal
S}=0}^{2S} g_{{\cal S}} \sum_{{\cal M}=-{\cal S}}^{\cal S}
|{\cal S},{\cal M} \rangle \langle {\cal S}, {\cal M}| \nonumber
\\
&+ \left(\boldsymbol{S}_1+\boldsymbol{S}_2\right)\cdot
\boldsymbol{B} \nonumber \\
&+ {d^2}
\frac{\boldsymbol{S}_1\cdot\boldsymbol{S}_2-3(\boldsymbol{n}\cdot\boldsymbol{S}_1)(\boldsymbol{n}\cdot\boldsymbol{S}_2)}{|\boldsymbol{r}_1-\boldsymbol{r}_2|^3},
\end{align}
where $\boldsymbol{r}_i=(x_i,y_i,z_i)$ and $\boldsymbol{S}(i)$
are positions and spin operators of the $i$-th atom
respectively. By $\boldsymbol{n}=\boldsymbol{r}/r$ we denote the
unit vector in the direction of
$\boldsymbol{r}=\boldsymbol{r}_1-\boldsymbol{r}_2=(x,y,z)$. Spin
operators
are represented by the spin $7\times 7$ matrices. To describe
the contact two-body interactions, which depend on the total spin
of both atoms, we use the molecular basis in the spin space:
\begin{equation}
|{\cal S},{\cal M}\rangle = \sum_{m_1,m_2}(S,m_1;S,m_2|{\cal
S},{\cal M})|S,m_1\rangle |S,m_2\rangle,
\end{equation} 
where $|{\cal S},{\cal M}\rangle$ is a two particle spin state
with the total spin ${\cal S}$ and the projection $\cal M$. The
symbol $(S,m_1;S,m_2|{\cal S},{\cal M})$ is the Clebsch-Gordan
coefficient and $|S,m_i\rangle$ describes the spin state of the
single atom of spin $S$ and its projection
$m_i$. Note that due to the symmetry the total spin of the
doublon, $\cal S$  can take only even values, ${\cal S} =0, 2,
4, 6$. The parameter $\kappa$ is: $\kappa = (\hbar \Omega_z)/(2
E_R)$. The strength of the contact interactions $g_{{\cal S}}$
is proportional to the s-wave
scattering length $a_{{\cal S}}$ in the given $\cal S$-channel:
\begin{equation}
g_{{\cal S}} = 8\pi k a_{{\cal S}}.
\end{equation} 
Finally the dimensionless dipole-dipole interaction strength
reads $d^2 = \mu_0/(4\pi)(g \mu_B)^2 (k^3/E_R)$, where $\mu_0$
is the magnetic constant.

The physical meaning of all the terms in the Hamiltonian
Eq.(\ref{Hamiltonian}) is straightforward. The first line
corresponds to the single particle energy, $H_0$, including both
kinetic and potential energies, while the second line describes
the contact interactions between particles,
$H_C$. In principle, the contact interactions can lead to a spin
dynamics. In this paper we assume that initially both atoms are
prepared in the ground state of
the lattice site in the presence of sufficiently strong magnetic
field to assure a ferromagnetic polarization of the system. The
initial spin state of the doublon is $|{\cal S},{\cal
M}\rangle=|6,6 \rangle $. In such a situation the spin dynamics
is driven,
at least initially, by the dipolar interactions only. The
external magnetic field, the third line of
Eq.(\ref{Hamiltonian}), describes the linear Zeeman shift of
single particles energies, $H_Z$. The magnetic field allows for
tuning of relative energies of the two atoms system. In this
paper we assume that the magnetic field
is align along the vertical
$z$-direction, perpendicular to the 2D lattice plane. Finally,
the last line of Eq.(\ref{Hamiltonian}) corresponds to the
energy of the dipole-dipole interactions of both atoms, $H_D$.
The total Hamiltonian is the sum
$H=H_0+H_C+H_Z+H_D$.

We examine the dipolar terms in more details stressing their
role on the spin dynamics. The elementary processes driven by 
the Hamiltonian $H_D$ can be casted into three classes.

1. Collisions not changing orbital angular momentum of the
doublon, $\Delta {\cal L}_z=0$, where ${\cal L}_z$ is the
$z$-component of the total orbital angular momentum of the pair.
These collisions are governed by the term whose spatial
dependence
has the form:
\begin{equation}
\label{h0}
h_{0}=\frac{d^2}{r^3}\left(1-\frac{3z^2}{r^2}\right),
\end{equation}
where and $r=|\boldsymbol{r}|$. Note that this term has to be
multiplied by the appropriate operator acting in the spinor
space:
\begin{equation}
\label{s0}
s_0= S_z(1)S_z(2)-{1 \over 4} \left(
S_{+}(1)S_{-}(2)+S_{-}(1)S_{+}(2)\right).
\end{equation} 
The $S_{\pm}(i) = S_{x}(i)\pm iS_{y}(i)$ operator rises ($+$) or
lowers ($-$) the magnetic quantum number of the $i$-th atom by
one quantum. The rising and lowering operators fulfill the
commutation relations for the spin, i.e $[S_{+}(k),S_{-}(j)]
=2iS_{z}(j)\delta_{kj}$. Note, that the
spinor term $s_0$ does not change the magnetization of the
doublon, $\Delta{\cal M}=0$. The dipolar interaction with
$\Delta {\cal L}_z=0$ are:
\begin{equation}
H_{d0}=h_0 \cdot s_0
\end{equation}
$H_{d0}$ evidently conserves the total angular momentum along
the $z$-axis, $\Delta{({\cal L}_z+{\cal M})}=0$. 
 
2. Collisions changing the $z$-component of the orbital angular
momentum of the doublon by one quantum, $\Delta {\cal L}_z= \pm
1$. These processes originate from the term of the following
spatial dependence:
\begin{equation}
\label{h1}
h_{\pm 1}=-\frac{3d^2}{2r^3}\frac{z(x \pm iy)}{r^2}.
\end{equation}
The upper sign corresponds to $\Delta {\cal L}_z=1$, while the
lower sign to $\Delta {\cal L}_z=-1$. This spatial term, $h_{\pm
1}$, has to be assisted by the corresponding spinor term:
\begin{equation}
\label{s1}
s_{\mp 1}=S_z(1)S_{\mp}(2)+S_{\mp}(1)S_{z}(2).
\end{equation}
The upper sign corresponds to processes with $\Delta {\cal M}=-1$ ,
while the lower sign to the processes with $\Delta {\cal M}=+1$.
The dipolar Hamiltonian describing processes with $\Delta {\cal L}_z=\pm 1$
 is a product of the spatial and the spinor terms:
\begin{equation}
H_{d1}=h_{+1}\cdot s_{-1}+h_{-1}\cdot s_{+1},
\end{equation} 
and clearly $\Delta{({\cal L}_z+{\cal M})}=0$. 

3. Collisions changing the $z$-component of the orbital angular
momentum of the doublon by two quanta, $\Delta {\cal L}_z= \pm
2$ . These processes originate from the term of the following
spatial dependence:
\begin{equation}
\label{h2}
h_{\pm 2}=-\frac{3 d^2}{4r^3}\frac{(x \pm i y)^2}{r^2},
\end{equation}
which is accompanied by the spin term which changes the spin
projection of the doublon by $\Delta {\cal M}=\mp 2$:
\begin{equation}
\label{s2}
s_{\mp 2}= S_{\mp}(1)S_{\mp}(2).
\end{equation}
The dipolar Hamiltonian describing processes with $\Delta {\cal
L}_z=\pm 2$ is a product of the spatial and the spinor terms:
\begin{equation}
H_{d2}=h_{+2}\cdot s_{-2}+h_{-2}\cdot s_{+2}.
\end{equation} 
Clearly the $z$-component of the total angular momentum of the
doublon $\Delta{({\cal L}_z+{\cal M})}=0$ is conserved. 

This way we decomposed the dipolar interactions into different
terms:
\begin{equation}
H_{D}=H_{d0}+H_{d1}+H_{d2},
\end{equation}
depending on the effect each term has on the orbital (or
equivalently the spin) dynamics of the atom pair.

\subsection{General considerations}
\label{general}

To start with we assume that the external potential near the
center of the site at ${\boldsymbol r}=0$ can be approximated by
the harmonic one. We keep only the lowest order terms in the
expansion of the potential in Eq.({\ref{Hamiltonian}}):
\begin{equation}
\label{V_har}
V_{tr}=V_0(x^2+y^2)+\kappa^2 z^2
\end{equation}
The potential is both harmonic and axially symmetric with
effective frequencies $\Omega_{\perp}=\omega_x=\omega_y=2
\sqrt{V_0}$ and $\Omega_z=2 \kappa$. The aspect ratio is
$\Omega_z/\Omega_{\perp}=\kappa/\sqrt{V_0}$.

In the following we assume that initially
both atoms are in the ground state of the external trap
in the two body $|{\cal S}=6,{\cal M}=6\rangle$ spin state, what
corresponds to a fully polarized system with the both atoms in $m=3$
state. The energy of the doublon is equall to a sum of single particle
and contact energies. The initial orbital state
$\Psi_{\rm ini} ({{\bf r}_1},{{\bf r}_2})=\langle {{\bf
r}_1},{{\bf r}_2}|{\rm ini}\rangle$ does not depend on the angles of
the vectors $\bf{r_1}$ and $\bf{r_2}$:
\begin{equation}
\Psi_{\rm ini} ({{\bf r}_1},{{\bf r}_2}) =
\phi_{00}(x_1,y_1)\varphi_0(z_1)\phi_{00}(x_2,y_2)\varphi_0(z_2),
\end{equation}
where $\phi_{00}(x,y) \varphi_0(z)$ is the ground state of the
harmonic trap. In order to stress that the axial
wavefunction, $\varphi_0(z)$, corresponds to a different trap frequency we
separated out the axial term.

To analyze further approximations we use the realistic
parameters related to experiments of \cite{Bruno_1}. We assume
that the optical lattice is created by the laser light of the
wavelength of $532$nm, which gives the unit of distance to be
$84,67$nm. For Chromium $^{52}$Cr ($M\simeq 87
\times 10^{-27}$kg) the recoil energy is $E_R/\hbar \simeq 2\pi
\times 13,5 {\rm kHz}$. A typical height of the potential barrier
is $V_0=25 E_R$ what corresponds to $V_0/\hbar = 2\pi \times 336
{\rm kHz} $ for which the characteristic trap frequency is
$\Omega_{\perp} = 10 E_R/\hbar= 2\pi
\times 135 {\rm kHz}$. We assume that the
axial frequency is $\Omega_z= 16 E_R/\hbar= 2\pi \times 218 {\rm
KHz}$.

These parameters lead to a characteristic size of the  system
of the order of the harmonic oscillator lengths $a_{\perp} = 38
{\rm nm}$ and $a_z= 30 {\rm nm}$. With two atoms per site the
atom density is about $n_0 \sim 2.8 \times 10^{15} {\rm
cm}^{-3}$. Corresponding contact interaction
in $S=6$ channel is $U_C/\hbar \sim 2\pi \times 18.9 {\rm kHz}$
while the dipolar interaction $E_{D}/\hbar \sim d^2 S \sim 2\pi
\times 0.37 {\rm kHz}$.

The dipolar energy is the smallest energy scale. Thus we can
ignore the term of dipolar interactions which does not change a
total magnetization of both colliding atoms, i.e. we neglect
processes with $\Delta {\cal M}_z=0$. The term leads only to
small, negligible corrections to
the contact interactions.

Clearly, the dipolar interactions leading to a spin dynamics
assisted by an inevitable transfer of atoms to excited states of
the trap are suppressed by the energy conservation because
$\hbar \Omega_{\perp} >> E_D$. The spin dynamics activates if
the energy gap is compensated by the Zeeman
energy. To shift the relative energies of $m=3$ and $m=2$ components
by the amount of $\hbar \Omega_{\perp}$ the magnetic field equal
to $B_{tr} = 48.2$mG is needed. Similarly, the Zeeman energy
compensating for the contact interactions is related to the
magnetic field $B_C \sim 6.75$mG,
while the dipolar energy
corresponds to $B_D \sim 25.7\mu$G. Extremely precise control
of bothL: the direction and the value of the magnetic field is necessary
to control the spin dynamics in the trap. This makes a great
challenge for experiments.
  
For the initial axially symmetric ground state $\Phi_{\rm
ini}=\langle{\rm r}_1,{\rm r}_2|{\rm ini}\rangle $ the processes
with $\Delta {\cal M}_z=-1$ and $\Delta {\cal M}_z=-2$ lead to
two different orbital states of the doublon. In the lowest order
of the perturbation in the dipolar
interactions, the angular dependence of the two atoms' relative
wavefunction (coupled to the ground state by dipolar
interactions) is determined by the angular dependence of the
dipolar terms, Eqs.(\ref{h1}), (\ref{h2}). The state excited due
to $\Delta {\cal M}_z= - 1$ dipolar process
has energy of the order of $\Omega_{\perp}+\Omega_{z}$ , while the
energy of the state reached by the system via the second
channel, $\Delta {\cal M}_z= - 2$, is $2 \Omega_{\perp}$. If
$(\Omega_z-\Omega_{\perp})$ is much larger than the dipolar energy
then by matching the resonance condition
one can activate selectively the first or the second channel (not
both of them). Therefore we will analyze the process with
$\Delta {\cal M}_z= - 1$ and $\Delta {\cal M}_z= -2$ separately.

Our general strategy is to find the lowest energy state which is
coupled by the dipole-dipole interactions to the $|{\rm ini}\rangle$
state as it follows directly from the structure of the dipolar
Hamiltonian. At this step we determine a single particle basis
which is needed to describe
all relevant two-body states. Then we diagonalize the full
Hamiltonian $H$ in the chosen basis and find eigenenergies and
eigenvectors of the doublon. At every resonance there exists an
eigenvector of the Hamiltonian which is an equal-amplitude
superposition of the initial and the final
state. The two states forming the superposition are coupled by
the dipolar interaction.

Alternatively, the physical insight can be gained with the help of
a perturbative analysis. The state determined by the structure
of the dipole-dipole term is not necessarily the eigenstate of the
single particle and the contact Hamiltonian. The goal is to
decompose this state in the
eigenstate basis of $H_0+H_Z+H_C$. A number of dipolar
resonances equals to a number of different eigenstates involved
in the decomposition. Each component has different eigenenergy
and therefore becomes resonant at a  different magnetic field.

To the `zero order' approximation the energy of a final state -
the state accessible by the dipolar interactions -- is given by
the sum of the single particle energies
(i.e. $\Omega_{\perp}+\Omega_{z}$ in the $\Delta {\cal M}_z=-1$
channel or $2\Omega_{\perp}$ in $\Delta {\cal M}_z=-2$
channel). The single particle energy gives an estimation of the
resonant magnetic field. However, anharmonicity and anisotropy
shift one-particle energies and the contact
interactions couple them. These effects
result in a  splitting of energies of different components of the
final states accessible via the  dipole-dipole interactions
i.e. in an
appearance of a fine resonant structure.

The strength of the resonant dipole-dipole interactions can be
characterized by the self energy of the initial $|{\rm
ini}\rangle$ state of energy $E_0(B)$:
\begin{equation}
\label{Es}
E_{sf}(B)=\sum_i \frac{|D_i|^2}
{\sqrt{\left(\frac{E_i(B)- E_0(B)}{2}\right)^2+|D_i|^2}}.
\end{equation}
$|\Psi_i\rangle$ and $E_i(B)$ are `bare' eigenstates and
eigenenergies of the two-body Hamiltonian without the  dipolar
interactions, $(H_0+H_C+H_Z)|\Psi_i\rangle=E_i(B)
|\Psi_i\rangle$, while $D_i=\langle{\rm ini}|H_D|\Psi_i\rangle$
is the dipole-dipole interaction matrix element. The self
energy contains a contribution from a virtual transfer of the
doublon from $|{\rm ini}\rangle$ to the intermediate
$|\Psi_i\rangle$ state. Evidently, this process contributes only
at the vicinity of the resonance, where:
\begin{equation}
E_i(B_{res})- E_0(B_{res}) \simeq D_i,
\end{equation}
The particular form of the self energy results from the fact,
that at every resonance we deal with a two level system, and
then the denominator in Eq.(\ref{Es}) gives the energy shift of
the $|{\rm ini}\rangle$ state.

\section{Dipolar processes with $\Delta {\cal M}_z=-1$}
\label{m1}

The spatial part of the lowest energy state of the doublon accessible
via  the dipole-dipole interactions in $\Delta {\cal M}_z=-1$ channel 
is $\sim z(x+iy)$, where $x,y,z$, are relative coordinates of two atoms.
This state is created by a two-particle
operator:
\begin{equation}
\label{vortex1}
v_1^{\dagger} = \frac{1}{2} \left( p^{\dagger}_z
\left(p^{\dagger}_x+ip^{\dagger}_y\right)
-s^{\dagger} \left( d^{\dagger}_{xz}+id^{\dagger}_{yz}\right)
\right),
\end{equation} 
acting on the particle vacuum. The bosonic operators in the
above formula create a particle in the one of the following 
states of the harmonic
oscillator: $s^{\dagger}$ -- in the
ground state $\phi_{00}(x,y)\varphi_0(z)$, $p^{\dagger}_x$, 
$p^{\dagger}_y$ --  in one of
the two $p$-shell states: $\phi_{10}(x,y)\varphi_0(z)$ or
$\phi_{01}(x,y)\varphi_0(z)$ respectively, $d^{\dagger}_{xz}$ and
$d^{\dagger}_{yz}$ --  in the $d$-shell states --
$\phi_{10}(x,y)\varphi_1(z)$ or $\phi_{01}(x,y)\varphi_1(z)$.
$\varphi_1(z)$  is the first excited state
in the axial direction. The operator $v_1^{\dagger}$ allows for
a compact representation of the `vortex' state with ${\cal
L}_z=1$. The vortex is created in a relative coordinate space of
the doublon. The first term in Eq.(\ref{vortex1}) describes the
state in which one atom takes one
excitation quanta in the axial direction, while the second
carries the vortex like excitation. The second term corresponds
to a process in which one atom experiences both types of
excitations while the second atom
is a spectator (at least as the
spatial wavefunction is considered).

The operator $v^{\dagger}_1$ defines only a spatial part of
${\cal L}_z=1$ vortex state:
\begin{equation}
\label{vor1}
|{\cal L}_1\rangle= v_1^{\dagger}|0\rangle.
\end{equation}
To account for a total state of the system including spin
degrees of freedom, the $|{\cal L}_1\rangle$ state has to be
multiplied by a totally symmetric spinor component with one atom
in $|m=3\rangle$ and the other in $|m=2\rangle$ spin states. The
final state of the system is:
\begin{equation}
\label{fin1}
|{\cal F}\rangle = \frac{1}{\sqrt{2}}\left(
|3\rangle|2\rangle+|2\rangle|3\rangle\right) |{\cal L}_1\rangle.
\end{equation} 

\begin{figure}[h]
\includegraphics{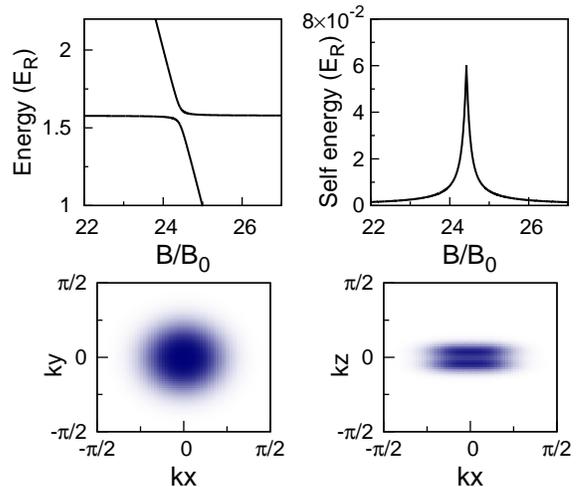}
\caption{Harmonic trap -- dipolar resonances with $\Delta {\cal M}_z=-1$. 
Left upper panel: the excitation  energy of the two particle system  in the harmonic trap as a function
of an external magnetic field.
The horizontal line at small $B$ corresponds to the $|{\rm ini}\rangle$ state with 
both atoms fully polarized $|3\rangle |3\rangle $. Only the lowest energy 
state $|{\cal L}_1\rangle$, the one which is   
coupled to the $|{\rm ini}\rangle$ state by $\Delta {\cal M}_z=-1$ dipole-dipole interaction term is shown.
Notice avoided crossing. Right upper panel: a single maximum
in the self-energy curve (right upper panel) signifying strong coupling of both involved states at resonant magnetic field.
Lower panel: reduced single particle density matrix component (the diagonal element with magnetization $m=2$) of the state $|{\cal L}_1\rangle$;
      left panel -- density cut at $xy$ plane for $z=1/\sqrt{\kappa}$, right panel -- density at $xz$ plane ($y=0$).       
}\label{resonance_m1}\end{figure}

The most important observation is
that the contact interactions vanish in the $|{\cal L}_1\rangle$
state because its wavefunction
vanishes identically if ${\bf r}_1={\bf r}_2$, i.e. $\langle
{\bf r}, {\bf r}|{\cal L}_1\rangle=0$. This state is not coupled
by the contact interaction to any other state of the doublon in the
trap. It is therefore the eigenstate of both the single particle
and the contact Hamiltonians. The 
dipole-dipole interactions couple efficiently the state to the
initial one if energies of the involved states do match. This
can happen if $E_{{\cal L}_1}=\Omega_{\perp}+\Omega_{z}+(2+3)B$
equals to the energy of the initial state
$E_{ini}=U_{33}+2(3B)$. The resonance condition is met at the 
magnetic field $B_{res}=(\Omega_{z}+\Omega_{\perp})-U_{33}$.
$U_{33}=(g_6/8) \sqrt{\Omega_{\perp}^2\Omega_z/\pi^3}$ is the
contact interaction of two particles in $m=3$ state and the
ground state of the harmonic potential.

The state is also robust if anharmonicity of a the lattice site
potential is taken into account. In our case the lattice
potential depends only on $x$ and $y$ coordinates. The
$z$-direction separates. This fact has very important
consequences: $d_{xz}$ and $d_{yz}$ states are separable.
Their single particle energies are sum of $p$-band energies. For
this reason the two components of the $|{\cal L}_1\rangle$
state, i.e. the one with two excited particle and the second
with only one excited particle, are degenerate. Anharmonicity of
the lattice potential results in an
equal shifts of both components of the $|{\cal L}_1\rangle$
state and does not lift their degeneracy. Moreover, this
conclusion holds even when axial symmetry of the external
potential is substituted by the $Z_2$ symmetry of the single
site of the lattice. $Z_2$ symmetry is sufficient to
assure the degeneracy of $p_x$ and $p_y$ orbitals.
The only effect of the lattice potential is to shift the
value of the resonant magnetic field.

The latter will not be true if $p_x$ and $p_y$ orbitals have
different energies. In such a case the $|{\cal L}_1\rangle$
state will be a superposition of two eigenstates of
different energies,
and each component could be separately addressed in a
dipolar process by an appropriate tuning
of the magnetic field. Two resonances will be observed in such a
case. 
 
Diagonalization of the full Hamiltonian in the oscillatory or
Wannier basis supports the above discussion. We find only one
low energy state coupled by the 
 dipole-dipole interactions to the
initial one. In Fig.\ref{resonance_m1} we show eigenenergies of
the two states. The avoided crossing,
signifying the dipolar coupling is clearly visible. In the lower
panel the single particle density in $m=2$ spin component is
shown ($xy$ and $xz$ cuts). Note the node in $z=0$ plane what is
typical for the first excited state in the
axial direction. The single particle density does no
resemble the structure typical for a single particle vortex
state. This is because the vortex is created in the relative
coordinates of two particles. In Fig.\ref{resonance_m1}
(upper right panel) the self energy of the initial state is
plotted. The hight of the resonant peak indicates the value of the dipolar matrix
element coupling the states. If the magnetic field is tuned to
the resonant value the system will perform Rabi oscillations
between the initial and the final state. The final state reached
via the dipole-dipole interactions is
highly entangled, Eq.(\ref{fin1}).
 
\section{Dipolar processes with $\Delta {\cal M}_z=-2$}
\label{m2}

\subsection{Harmonic trap}

In this section we examine collisions changing magnetization of the system
by two quanta, $\Delta {\cal M}_z= -2$. The two interacting
particles, trapped in a harmonic potential, go simultaneously from
$|3\rangle |3\rangle$ to $|2\rangle |2\rangle$ spin state.
Spatial part of the relative wavefunction in the lowest
accessible final state must be proportional to $\sim (x + i
y)^2$ in order to compensate for a decrease of the spin of the
doublon by two quanta. This is the wavefunction of
a doubly charged vortex in the relative coordinates space. This
simple state (normalized) if written in the harmonic basis looks
quite complicated, so it is convenient to describe it using the
second quantization picture. The `vortex' state is created by
the two-body operator
$v^{\dagger}_2$:
\begin{eqnarray}
\label{v2}
v^{\dagger}_2 &=& \frac{1}{2 \sqrt{2}} \left( s^{\dagger}\left(
d^{\dagger}_{xx}+i
\sqrt{2}d^{\dagger}_{xy}-d^{\dagger}_{yy}\right)
\right. \\ \nonumber
&&\left.- \frac{1}{\sqrt{2}} \left( (p^{\dagger}_{x})^2 +i 2
p^{\dagger}_{x}p^{\dagger}_{y}- (p^{\dagger}_{y})^2 \right)
\right).
\end{eqnarray} 
In addition to defined previously bosonic operators creating
a particle in $s$ and $p$ shells we define the $d$ shell operators:
$d^{\dagger}_{xx}$, $d^{\dagger}_{yy}$, and $d^{\dagger}_{xy}$
creating a particle in the second excited state of a harmonic
oscillator potential, i.e.
$\phi_{20}(x,y)\varphi_0(z)$, $\phi_{02}(x,y)\varphi_z(0)$, and
$\phi_{11}(x,y)\varphi_0(z)$ respectively.

The first line describes the state with one particle in the
ground $s$ state and the second particle in the excited
$d$-shell state with ${\cal L}_z=2$. The second line corresponds
to a state with two particles occupying the $p_x+ip_y$ vortex.
Let us note that introduced here states,
abbreviated by $s$, $p_x$, $p_y$, $d_{xx}$, $d_{yy}$, and
$d_{xy}$, span a basis of the orbital states which is sufficient to
describe the lowest dipolar excitations in $\Delta {\cal M}_z=-2$
channel.

The orbital basis has to be completed by the spin states. These
are the initial $|3\rangle |3\rangle$ and the final $|2\rangle
|2\rangle$ states. The further spin dynamics due to the spin
dependent contact interactions  can subsequently induce a
transition from $|2\rangle |2\rangle$ to
$(|1\rangle |3\rangle+|3\rangle |1\rangle)/\sqrt{2}$ state.
However, the process can be suppressed due to the energy
mismatch if $m=2$ state is shifted by light. Therefore for a
moment we neglect contact spin-changing collisions.

We diagonalize the total Hamiltonian Eq.(\ref{Hamiltonian}) in
the basis of the two particle states spanned by the introduced
above single particle $s$, $p$ and $d$ orbitals in the initial
$m=3$ and the final $m=2$ spin component for different values of
an external magnetic field.

However to gain an insight into physical processes involved we
shall describe the results using perturbative approach with a
dipole-dipole interaction energy as a small parameter. First we
diagonalize the single particle and contact Hamiltonian
$H_0+H_Z+H_C$. As expected, the state
$|{\cal L}_2\rangle$, defined as:
\begin{equation}
\label{vortex}
|{\cal L}_2\rangle = v^{\dagger}_2|0\rangle,
\end{equation}
is the eigenstate of $H_0+H_Z+H_C$. Moreover, the contact energy
vanishes in this state. This is a unique feature of the harmonic
potential and has some important consequences on the spin
dynamics as it will be shown later. The energy of this state
$E_{{\cal L}_2}=2\Omega_{\perp}+2 (2B)$
equals to the energy of the initial state $E_{ini}=U_{33}+2(3B)$
at the magnetic field $B_{res}=\Omega_{\perp}-U_{33}/2$.
$U_{33}=(g_6/8) \sqrt{\Omega_{\perp}^2\Omega_z/ \pi^3}$ is the
contact interaction of two particles in $m=3$ state and the
ground state of the harmonic potential.

\begin{figure}[h]
\includegraphics{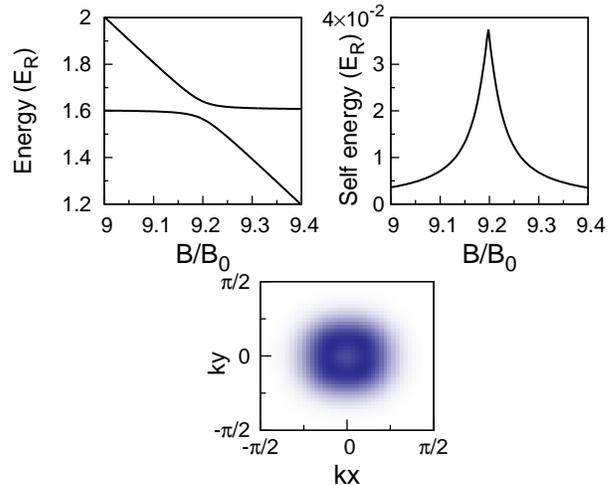}
\caption{ Harmonic trap -- dipolar resonances with $\Delta {\cal M}_z=-2$. 
Left upper panel: excitation  energy of the two particle system  in the harmonic trap as a function
of an  external magnetic field.
The horizontal line at small $B$ corresponds to the $|{\rm ini}\rangle$ state with 
both atoms fully polarized $|3\rangle |3\rangle $. 
Notice avoided crossing. Right upper panel: a single maximum
in the self-energy curve signifying strong coupling of both involved states at the 
resonant magnetic field.
Lower panel: reduced single particle density (in $xy$ plane) of $|{\cal L}_2\rangle$ state coupled to the $|{\rm ini}\rangle$ state.
}\label{harmonic}
\end{figure}

Evidently, by the construction, $|{\cal L}_2\rangle$ is coupled
to the initial state: $D=\langle {\cal L}_2|h_{+2}|{\rm
ini}\rangle \neq 0$. By diagonalization of the full Hamiltonian
with the dipole-dipole term included, we indeed find that
$|{\cal L}_2\rangle$ state is the only one
coupled to $|{\rm ini}\rangle$ state by the dipole-dipole
interactions. The two coupled states, $|\rm ini \rangle$ and
$|{\cal L}_2\rangle$ form a two level system.
Its eigenenergies depend on the value of the external
magnetic field. They are shown in
Fig.\ref{harmonic}. The visible  avoided crossing
corresponds to the resonance. Obviously there are other
resonant states, not included in our basis, which are coupled by
the dipolar interactions to the initial state. This coupling takes
place, however at higher magnetic fields
which we do not consider here.

In the upper right panel of Fig.\ref{harmonic} we show the self
energy of the $|{\rm ini}\rangle$ as a function of the magnetic
field. The self energy reaches maximum at the resonance. The
magnitude of $E_{sf}$ at the resonance is equal to $|D|$.
Similarly, a width of the resonance is also
equal to $|D|$.

Weakness of the dipolar coupling allows for a precise tuning of
the two atom system. Initial population of the $m=3$ state will
oscillate periodically between two coupled states with a Rabi
frequency proportional to the dipolar energy $D$. At half of the
period all the population will be
transferred to the $m=2$ state, $|{\cal L}_2\rangle$. The single
particle density corresponding to this state is shown in the
lower panel of Fig.\ref{harmonic}. Note a shallow minimum at
the trap center. The vortex $|{\cal L}_2\rangle$ state is a
vortex in the relative coordinate space.
This is not a vortex in the reduced one-particle space. 

Finally we return to the observation that the contact
interaction in $|{\cal L}_2\rangle|2\rangle |2\rangle$ state
vanishes, i.e. $U_{22}=g_{22} \langle {\cal L}_2|\delta({\bf
r}_1-{\bf r}_2) |{\cal L}_2\rangle = 0$, where
$g_{22}=(6g_6+5g_4)/11 $. This is because the two body
wavefunction of this state vanishes if ${\bf r}_1={\bf r}_2$.
The contact interactions leading to the transfer of one atom to
$m=3$ and the second one to $m=1$ state,
$U_{22,31}=g_{22,31}\langle {\cal L}_2|\delta({\bf r}_1-{\bf
r}_2) | {\cal L}_2\rangle = 0$, also vanishes, where
$g_{22,31}=\sqrt{30}/11(g_6-g_4)$. Therefore, in the harmonic
axially symmetric trap the spin dynamics of the $|{\cal
L}_2\rangle |2\rangle |2\rangle$ state due to the contact spin
changing collisions is forbidden. There is no need to detune
$m=2$ state with respect to $m=3$ and $m=1$
states to prevent contact spin changing collisions. This fact is
in a contradiction with experiments with Chromium atoms in an
optical lattice \cite{Bruno_1} and clearly indicates that
the harmonic approximation does not describe correctly
the spin dynamics in the optical lattices.

\subsection{Anharmonic trap}   

If in the expansion of a the lattice site potential we keep the
higher order terms then we get a potential which is neither
harmonic nor axially symmetric:
\begin{equation}
V({\bf r})=V_0 \rho^2-\frac{1}{3}V_0 \rho^4+\kappa^2 z^2
+\frac{2}{3}V_0 x^2y^2,
\end{equation}
where $\rho^2=x^2+y^2$. The second term $\sim
\rho^4$ introduces anharmonicity while the last term brakes the
axial symmetry. To separate effects of anharmonicity from
effects of anisotropy we neglect the last term and consider
the model anharmonic potential
\begin{equation}
\label{V_anhar}
V({\bf r})=V_0 \rho^2-\frac{1}{3}V_0 \rho^4+\kappa^2 z^2.
\end{equation}
The combined effect of anharmonicity and anisotropy of the
square lattice site will be discussed in the subsequent section.

A measure of anharmonicity is given by a ratio of the anharmonic 
to the harmonic energy, $\gamma=\frac{1}{3}\rho_0^2$, 
at the typical distance $\rho_0 = 1/\sqrt{\Omega_{\perp}}$.
Therefore the anharmonicity coefficient is:
\begin{equation}
\gamma=\frac{1}{6}\frac{1}{\sqrt{V_0}}.
\end{equation}
Obviously anharmoniciy $\gamma$ is small for deep optical
lattices.
The energy levels of an anharmonic potential are not equally
spaced. For this reason the state $|{\cal L}_2\rangle$ is no
longer the eigenstate of the Hamiltonian $H_0+H_Z+H_C$.

\begin{figure}[h]
\includegraphics{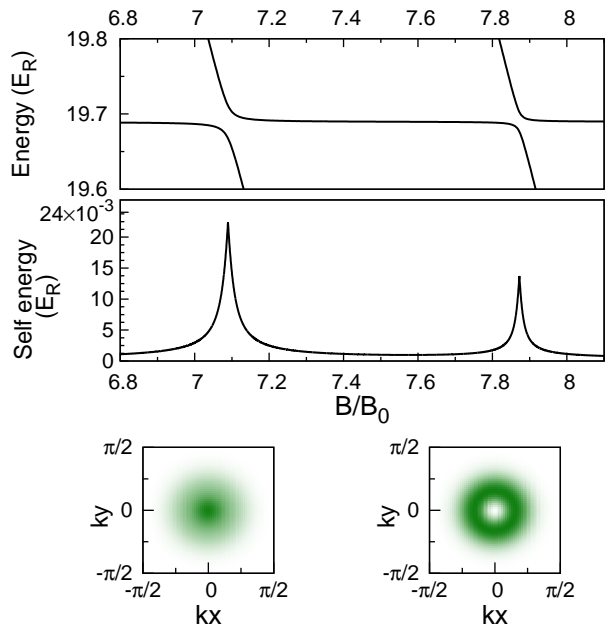}
\caption{ Anharmonic trap -- dipolar resonances with $\Delta {\cal M}_z=-2$.
Upper panel: excitation  energy of the two particle system  in the anharmonic trap as a function
of an external magnetic field. The horizontal line at small $B$ corresponds to the $|{\rm ini}\rangle$ state with 
both atoms fully polarized $|3\rangle |3\rangle $.    
Notice two avoided crossings. Middle panel: two maxima
in the self-energy curve signifying strong coupling  at the resonant magnetic field.
Lower panel: reduced single particle densities (in $xy$ plane)
of the final states accessible via $\Delta {\cal M}_z=-2$ 
resonant dipolar processes. The densities correspond to the two resonant values
 of the  magnetic field as indicated in the middle panel.  
         }\label{fig:anharmonic}
\end{figure}

To construct two-body eigenstates we find numerically all
$s$, $p$ and $d$ eigenstates of the
anharmonic potential. Then the full two-body Hamiltonian is
diagonalized in this basis. There are two states
$|v_{a_1}\rangle$ or $|v_{a_2}\rangle$ which are coupled to
the initial one at two different magnetic fields. The resonances
are visible as avoided crossings in Fig.\ref{fig:anharmonic}. At
each resonance the system is effectively a two level one.

The single particle densities of the two resonantly coupled
states for $V_0=25 E_R$ are presented in the lowest panel of
Fig.\ref{fig:anharmonic}. Let us note, that one particle density
of the $|v_{a_2}\rangle$ state resembles a singly quantized
vortex, but does not reach the zero value
at the center. In the middle panel of Fig.\ref{fig:anharmonic}
we show the self energy of the $|{\rm ini}\rangle$ state as a function
of the magnetic field. The self energy reaches a maximum at
every resonance.

The presented here numerical results can be understood with the
help of a perturbative approach with the dipolar energy being a
small parameter. The state $|{\cal L}_2\rangle$ can be viewed as
the superposition of the two states:
\begin{equation}
|{\cal L}_2\rangle = \frac{1}{\sqrt{2}}\left(|v_d\rangle
-|v_p\rangle\right),
\end{equation}
where  the $d$-shell state $|v_d\rangle$ is:
\begin{equation}
\label{eqn:v_d}
|v_d\rangle = \frac{1}{2}\left( s^{\dagger}\left(
d^{\dagger}_{xx}+i \sqrt{2}d^{\dagger}_{xy}-d^{\dagger}_{yy}
\right)\right)|0\rangle,
\end{equation}
and the $p$-shell state $|v_p\rangle$ has the following form:
\begin{equation}
|v_p\rangle = \frac{1}{2 \sqrt{2}}\left( (p^{\dagger}_{x})^2+i 2
p^{\dagger}_{x}p^{\dagger}_{y}-(p^{\dagger}_{y})^2\right)|0\rangle,
\label{eqn:v_p}
\end{equation}
In Eq.(\ref{eqn:v_d}) and Eq.(\ref{eqn:v_p}) the operators
$s$, $p$ , $d$ are not the same as those in the harmonic case
however we use the same notation. Now the operators create a particle 
in $s$ , $p$ , $d$ states of the anharmonic potential.

The two resonant states are the eigenstates of the z-component of the
orbital angular momentum of two particles. $|v_d\rangle$ is the
state with one particle in the ground and the second in the
$d$-state with ${\cal L}_z=2$. $|v_p\rangle$ is the
state with two particles occupying the same
$p$ orbital with ${\cal L}_z=1$. Evidently the single particle
density of the second state is typical for the singly charged
vortex state in a harmonic trap.

In the harmonic trap the two states have the same energies. Their
relative shift introduced by anhamonicity, $\delta E$, can be
estimated using the perturbative approach with $\gamma$ being
the small parameter:
\begin{equation}
\delta E = 2 \gamma \Omega_{\perp}={2 \over 3}.
\end{equation}
The energy of $|v_d\rangle$ state is smaller then the energy of
$|v_p\rangle$ state by the $2/3$ of the recoil energy. The shift
does not depend on the $V_0$.

These states are dressed by the contact interaction whose
strength is:
\begin{equation}
U_{22}=\frac{g_{22}}{16 \pi^{3/2}} \sqrt{\Omega_z}
\Omega_{\perp},
\end{equation}
where $g_{22}=(6g_6+5g_4)/11$. 
The dressed eigenvectors of this effective two
level system are solutions of a simple eigenvalue problem:
\begin{equation}
\label{2by2}
\begin{pmatrix}
-\delta E & U_{22}\\
U_{22}          & 0
\end{pmatrix}
\begin{pmatrix}
\alpha_{a}\\
\beta_{a}
\end{pmatrix}
=
\varepsilon_{a}
\begin{pmatrix}
\alpha_{a}\\
\beta_{a}
\end{pmatrix}.
\end{equation}
The two eigenstates
$|v_{a_1}\rangle=\alpha_{a}|v_d\rangle-\beta_{a}|v_p\rangle$ and
$|v_{a_2}\rangle=\beta_{a}|v_d\rangle+\alpha_{a}|v_p\rangle$ are
superpositions of the two coupled states $|v_d\rangle$ and
$|v_p\rangle$.

If the contact interaction significantly exceeds the energy
splitting $U_{22} >> \delta E/2$, i.e. for:
\begin{equation}
\label{recovery_har}
V_0 >> \frac{256}{9}\frac{\pi^3}{(g_{22})^2 \Omega_z},
\end{equation}   
then the harmonicity is recovered in a deep optical lattice
in the strong coupling limit. 

Note the crucial role of the axial trapping frequency. In a prolate
geometry the harmonicity is recovered for much deeper traps than in
a case of an oblate geometry. For $\Omega_z =16$ as used here,
the anharmonicity of the optical
lattices becomes negligible if $V_0>>29 E_R$. In this limit the
eigenvectors of the Hamiltonian are $|v_{a_1}\rangle \simeq
(|v_d\rangle-|v_p\rangle)/\sqrt{2}$ and $|v_{a_2}\rangle \simeq
(|v_d\rangle+|v_p\rangle)/\sqrt{2}$. Note, that the eigenstate
$|v_{a_1}\rangle$ corresponds to
relative excitations and is the same as the state
reached by the dipolar interactions in the harmonic trap,
$|v_{a_1}\rangle \simeq |{\cal L}_2\rangle$. The second one,
$|v_{a_2}\rangle$, decouples from the initial $|{\rm
ini}\rangle$ state as the corresponding dipolar matrix
element vanishes. This states corresponds to the center of mass
excitation. Obviously excitation of the center of mass is not
possible in the harmonic trap by any two-body interactions. We
checked that even for deep optical lattices, i.e. $V_0= 40 E_R$,
the effect of anharmonicity cannot
be neglected.

\begin{figure}[h]
\includegraphics[scale=.6]{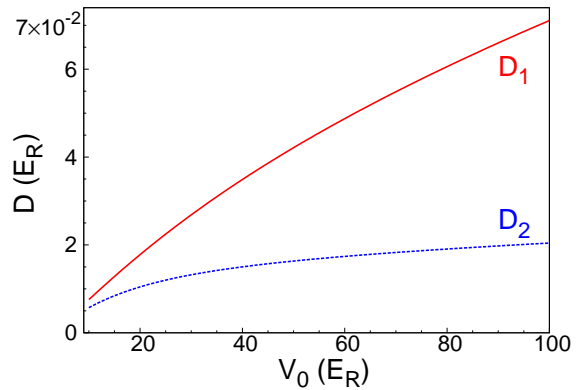}
\caption{Dipolar matrix elments $D_1=|\langle v_{a_1}|H_D|{\rm ini}\rangle|$ and
$D_2=|\langle v_{a_2}|H_D|{\rm ini}\rangle|$ for an axially symmeric anharmonic potential.
With increasing barrier height the coupling to the state with no center
of mass excitation dominates $D_1 \sim \sqrt{V_0}$ and the harmonic limit is recovered}\label{anharm_D}
\end{figure}

The way the harmonicity 
is recovered is quite interesting. First, the strong coupling results in a 
very large splitting of resonant energies with respect to the harmonic case.
Separation of both resonances increases with a lattice height. Moreover
the dipolar matrix element 
$D_1=\langle v_{a_1}|H_D|{\rm ini}\rangle$ grows 
with the barrier height $D_1\sim V_0^{1/2}$, while 
the coupling to the second resonant state $D_2=\langle v_{a_2}|H_D|{\rm ini}\rangle$
is small and independent  of $V_0$ ($\sim \delta E$), see Fig.\ref{anharm_D}.
This way for very high lattices $D_2/D_1$ becomes very small and
only one resonant state is resonantly coupled via the dipole-dipole intaractions.

The other extreme is shallow optical lattice. The contact interactions
become negligible, $U_{22}<< \delta
E/2$. The eigenstates of the interacting systems can be
approximated by the $|v_{a_1}\rangle \simeq |v_d\rangle$ and
$|v_{a_2}\rangle \simeq |v_p\rangle$. The one-particle density
of the latter state is characteristic for a
singly quantized vortex - it is therefore a perfect candidate to
observe the EdH effect. The case shown in
Fig.\ref{fig:anharmonic} corresponds to a crossover situation
$\delta E \sim U_{22}$.
 
Till now we did not take into account the spin dynamics due to
the contact interaction in the final $|2\rangle |2\rangle$ spin
space of the doublon. This simplification is justified if
coupling due to contact interactions is `turned
off' by a relative light shift of energies of $m=3$ and $m=2$
states. If it is not the case then the contact interactions
following the dipolar process can further transfer the doublon
to $|3\rangle |1\rangle$ spin state.

Somewhat oversimplified picture of $|2\rangle |2\rangle$ to $|3\rangle
|1\rangle$ coupling is based on the observation that the
eigenstates $|v_{a_1}\rangle$ and $|v_{a_2}\rangle$ of the two body
contact Hamiltonian have their analogons of the same eigenenergies
and the same wavefunctions in $|1\rangle
|3\rangle$ subspace. If so, then the state
$|v_{a_i}\rangle|2\rangle |2\rangle$ ($i=1,2$) is coupled to
$|v_{a_i}\rangle(|1\rangle |3\rangle+|3\rangle
|1\rangle)/\sqrt{2}$ state due to the contact term $g_{22,31}
\langle v_{a_i}|\delta({\bf r}_1-{\bf r}_2)|v_{a_i}\rangle$.
Then the form
of the two-body wavefunction in $|3\rangle |1\rangle$ space is
inherited from the $|2\rangle |2\rangle$ subspace.

Situation is more complicated however. The contact interactions
are different in different spin subspaces. They are proportional
to $g_{22}=(6g_6+5g_4)/11$ or to $g_{31}=(5g_6+6g_4)/11$ in
$|2\rangle |2\rangle$ or in $|3\rangle |1\rangle$ spin subspace
respectively. For this reason the
vectors diagonalizing contact interactions in
$|3\rangle|1\rangle$ space are different then $|v_{a_1}\rangle$
and $|v_{a_2}\rangle$ diagonalizing the contact interactions in
the $|2\rangle |2\rangle$ space. Their eigenenergies are also
different. Note however that difference in the
contact interaction strength is
$g_{22}-g_{31}=g_{22,31}/\sqrt{30}$, i.e. by factor
$\/\sqrt{30}$ smaller than the term coupling both spin
subspaces $g_{22,31}$. For this reason, the simplified
discussion is not totally wrong.

Exact treatment requires taking into account two pairs of
vectors from different subspaces. In order to find the states of
the doublon dressed by the contact interactions one has to
diagonalize the $4\times 4$ matrix in the basis of $|v_d\rangle$
and $|v_p\rangle$ states in the two involved spin subspaces. 
The splitting of energies of the four eigenstates is determined 
by the contact interactions. The eigenstates are superpositions 
of different spin states of the doublon. Each of them can 
be reached in the dipolar relaxation process if the external 
magnetic field is tuned to the given resonant value.

\begin{figure}[h]
\label{anharmonic}
\includegraphics{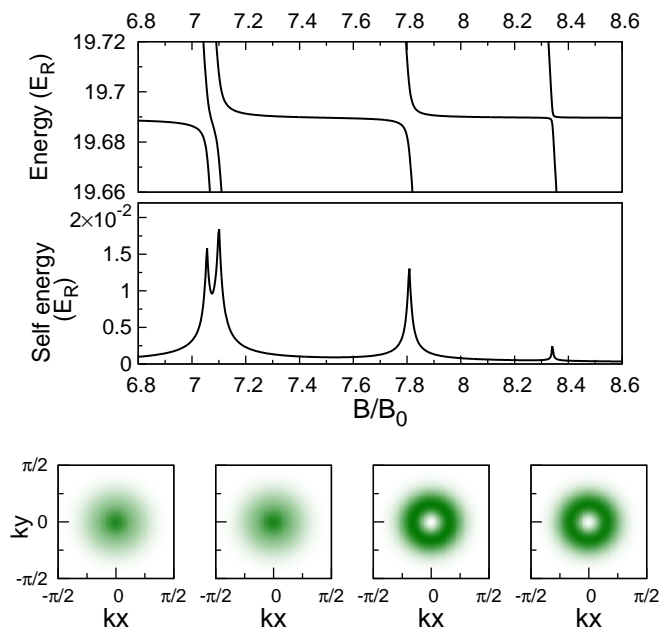}

\caption{Anharmonic trap -- dipolar resonances with $\Delta {\cal M}_z=-2$
when the contact interaction coupling  $|2\rangle |2\rangle$ and  $|3\rangle |1\rangle$ 
spin states is included (compare Fig.\ref{fig:anharmonic} where this coupling
was omitted).
Upper panel: 
excitation  energy of the two particle system  in the anharmonic trap as a function
of an external magnetic field (upper panel). The horizontal line at small $B$ 
corresponds to the $|{\rm ini}\rangle$ state with 
both atoms fully polarized $|3\rangle |3\rangle $.   
Notice four avoided crossings. Middle panel: four maxima
in the self-energy curve signifying strong coupling
at resonant magnetic field.
Lower panel: reduced single particle densities (in $xy$ plane) of the final states
accessible via $\Delta {\cal M}_z=-2$ resonant dipolar processes
and the spin dependent contact interactions
corresponding to the four resonant values 
of the  magnetic fields as indicated in the middle panel.  
}\label{fig:anh_31}
\end{figure}

In Fig.\ref{fig:anh_31} the four resonances are clearly visible
for $V_0=25 E_R$. The single particle densities corresponding to
the every resonant state of the doublon are shown in the lowest
panel. Let us observe first that the parameters we choose are in
the crossover region. The
anharmonic shift of energies, although does not dominate the
contact interactions, is still important. The low magnetic
field resonant state, in both spin subspaces, is dominated by
the $d$-shell $|v_d\rangle$ orbital with one particle in the
trap ground state (low panel of
Fig.\ref{fig:anh_31}). That is why the single particle density
is large in the trap center. The higher field resonant state is
dominated by the $p$-shell excitation. This is the reason of the
density minimum at the center.

A small anharmonicity of the binding potential significantly
modifies the spin dynamics. In the harmonic trap the two atoms
oscillate between $|{\rm ini}\rangle |3\rangle |3\rangle$ and
$|{\cal L}_2\rangle |2\rangle |2\rangle$ state at the resonant
magnetic field $B_{res}$. The $|{\cal
L}_2 \rangle$ state is characterized by the orbital angular
momentum ${\cal L}_z=2$ with no center of mass excitations. For
a realistic lattice depth this single resonance splits into
four resonances. The wavefunctions of the doublon are,
in all four cases, eigenstates of the
${\cal L}_z$ component of the orbital angular momentum (with the
center of mass excitation). The states are entangled. This
entanglement can be usefull, particularly after applying a
Stern-Gerlach magnetic field separating spatially different
entangled magnetic components of the
doublon.  

\subsection{Single site of a square lattice}
 
Finally we are ready to discuss a realistic situation of two
atoms in a single site of the square optical lattice where both:
the effect of anharmonicity and anisotropy are present
simultaneously. Again, in order to find final states of the doublon
which are accessible via dipolar
interactions starting from the initial $|{\rm ini}\rangle |3\rangle
|3\rangle$ state we have to find the eigenstates of the single
particle and contact Hamiltonian $H_0+H_Z+H_C$. To this end we
have to adjust the single particle basis to the lattice case.
Instead of oscillatory wavefunctions we
use the Wannier states (except of the $z$-component part which
is the ground state wavefunction of the harmonic oscillator in
the z-direction. To simplify a notation we omit the z-dependence
of the basis wavefunction - but we keep it in our calculations).
If we limit our basis to the
three lowest bands in the lattice then the best approximation to
the vortex state is the state created by the operator
$v_2^{\dagger}$ as defined in Eq.(\ref{v2}). The operators
$s^{\dagger}$, $p_x^{\dagger}$, $p_y^{\dagger}$,
$d_{xx}^{\dagger}$, $d_{yy}^{\dagger}$, $d_{xy}^{\dagger}$
have slightly different meaning now. They generate a particle in
the Wannier states ${\cal W}_l(x)$ rather than in the harmonic
oscillator states. For example $s^{\dagger}$ creates particle in
the $\Psi_{00}={\cal W}_0(x){\cal W}_0(y)$ state,
$d_{xx}^{\dagger}$ in the $\Psi_{20}={\cal
W}_2(x){\cal W}_0(y)$ state, or $d_{xy}^{\dagger}$ in the
$\Psi_{11}={\cal W}_1(x){\cal W}_1(y)$ state, etc.

It is convenient to rearrange the terms in the expression for
the vortex state $|{\cal L}_2\rangle$ according to their single
particle energies in the lattice, i.e:
\begin{equation}
|{\cal L}_2\rangle = |{\rm v}_1\rangle + |{\rm v}_2\rangle +
|{\rm v}_3\rangle,
\end{equation} 
where:
\begin{eqnarray}
|{\rm v}_1\rangle & = &
\frac{1}{2\sqrt{2}}\left(s^{\dagger}\left(d_{xx}^{\dagger}-d_{yy}^{\dagger}\right)\right)|0\rangle,
\\
|{\rm v}_2\rangle & = &
-\frac{1}{4}\left(p_x^{\dagger}p_{x}^{\dagger}-p_y^{\dagger}p_{y}^{\dagger}\right)|0\rangle,
\\
|{\rm v}_3\rangle & = &
\frac{i}{2}\left(s^{\dagger}d_{xy}^{\dagger}-p_x^{\dagger}p_{y}^{\dagger}\right)|0\rangle.
\end{eqnarray}
The single particle contribution to the energies of these states
are $E_1=E_s+E_d$, $E_2=2E_p$, and $E_3=E_2$ respectively, where
$E_i^{(1D)}=\langle {\cal W}_i|H_0|{\cal W}_i\rangle$ is the
mean energy of particle in the given Wannier state. The single
particle energies of the last two
states are equal because of the $Z_2$ symmetry of the problem.
Anharmonicity is responsible for the inequality $E_1 < E_2$.

The goal is to find eigenstates of the contact Hamiltonian and
decompose $|{\cal L}_2 \rangle$ state in the basis of doublon
eigenstates. Note, that the two-body wavefunction $\langle {\bf
r}_1, {\bf r}_2 |{\rm v}_3\rangle$ vanishes identically, if ${\bf
r}_1 = {\bf r}_2$, so does the
contact interaction in this state as well as all possible
contact interactions coupling this state to the other states.
The state $|{\rm v}_3\rangle$ is therefore the eigenstate of the
two-body Hamiltonian $H_0+H_Z+H_C$.

On the contrary contact interactions couple the vectors $|{\rm
v}_1\rangle$ and $|{\rm v}_2\rangle$. The states $|{\rm
v}_1\rangle$ and $|{\rm v}_2\rangle$ are analogous to
$|v_d\rangle$ and $|v_p\rangle$ states discussed in the previous
section. Their two linear superpositions
diagonalize the contact Hamiltonian. The two eigenvectors can be
found numerically.

\begin{figure}[h]
\includegraphics{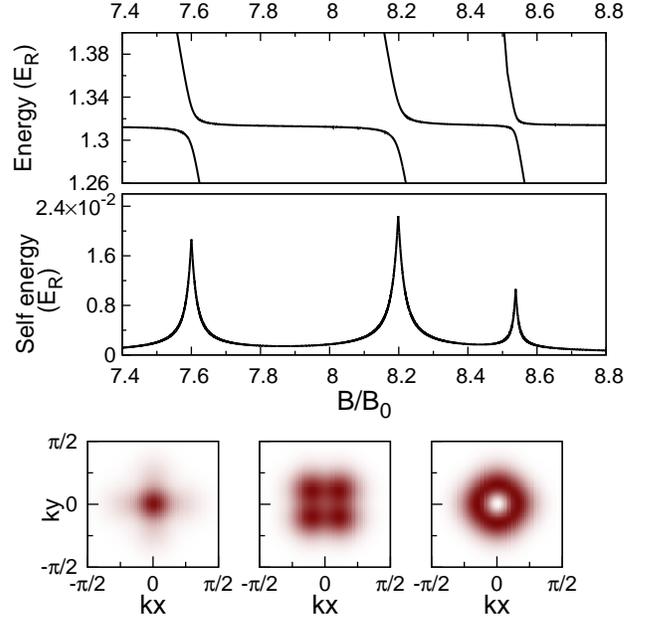}
\caption{Single site of a square lattice -- dipolar resonances with $\Delta {\cal M}_z=-2$ 
Upper panel:
excitation  energy of the two particle system  in a single
lattice site as a function of an external magnetic field.
The horizontal line at small $B $corresponds to the $|{\rm ini}\rangle$ state with 
both atoms fully polarized $|3\rangle |3\rangle $. 
Notice three avoided crossing. Middle panel: three maxima
in the self-energy curve signifying strong coupling  at the resonant magnetic field.
Lower panel: reduced single particle densities (in $xy$ plane)
of the final states accessible via $\Delta {\cal M}_z=-2$ 
resonant dipolar processes  corresponding to the three resonant values
 of the  magnetic fields as indicated in the middle panel. 
 }\label{Wannier3}
\end{figure}

In the absence of coupling to the $|3\rangle|1\rangle$ subspace,
there are three low energy states which can be accessed due to
the dipole-dipole interactions. The three two-body eigenenergies
(meassured from the energy of the initial state) are shown in
Fig.\ref{Wannier3}. Avoided
crossings signifying dipole coupling are clearly visible. In the
lower panel the single particle densities of the states coupled
to the initial one are shown, while in the middle panel the self
energy of the initial state is presented. The self energy
depicts position, strength and width
of every resonance. Note, that for axially symmetric harmonic
trap, all three resonances will coincide forming ${\cal L}_z=2$
vortex state in the relative coordinates space of the two atoms.

\begin{figure}[h]
\includegraphics{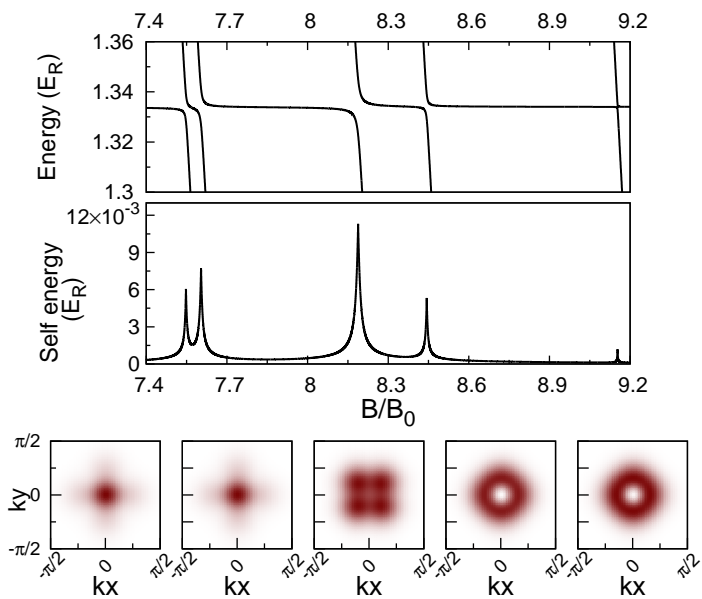}
\caption{Single site of a square lattice -- dipolar resonances with $\Delta {\cal M}_z=-2$ 
when the contact interaction
coupling  $|2\rangle |2\rangle$ and  $|3\rangle |1\rangle$ 
spin states is included (compare Fig.\ref{Wannier3} where this coupling
was omitted).
Upper panel: the excitation  energy of the two particle system  in a single lattice site as a function
of external magnetic field. The horizontal line at small $B$ corresponds to the $|{\rm ini}\rangle$ state with 
both atoms fully polarized $|3\rangle |3\rangle $. 
Notice five avoided crossings. Middle panel:  five maxima
in the self-energy curve signifying strong coupling  at the resonant magnetic field.
Lower panel: reduced single particle densities (in $xy$ plane) of the final states
accessible via $\Delta {\cal M}_z=-2$ resonant dipolar processes
and the spin dependent contact interactions
corresponding to the five resonant values 
of the  magnetic fields as indicated in the middle panel. 
}\label{Wannier5}
\end{figure}

If we account for the contact interactions coupling $|{\rm
v}_1\rangle$ and $|{\rm v}_2\rangle$ states in
$|2\rangle|2\rangle$ spin subspace to analogous states in
$|3\rangle|1\rangle$ subspace the additional channels of the
spin dynamics do appear. The situation is analogous to the one
discussed in the case of anharmonic trap. Again, the $4\times 4$
eigenvalue problem leads to four different eigenvectors with
entangled spin and spatial wavefunctions. Together with $|{\rm
v}_3\rangle$ state there are five low energy dipolar states
resonantly coupled to the initial one
in the single site of the optical lattice. The state $|{\rm v}_3\rangle$
is not shifted by contact interactions, `lives' entirely in the
$|2\rangle|2\rangle$ spin subspace. All the resonances 
obtained by numerical diagonalization of the full Hamiltonian
in the Wannier basis are
visible in Fig.\ref{Wannier5}. 
The single particle densities of the states
accessible by the dipole-dipole interactions (summed over
final spin projection) are shown in the lower panel.

\section{Comparison with the experiment}
\label{experiment}

\begin{figure*}[t!]
\begin{center}
\includegraphics[scale=1]{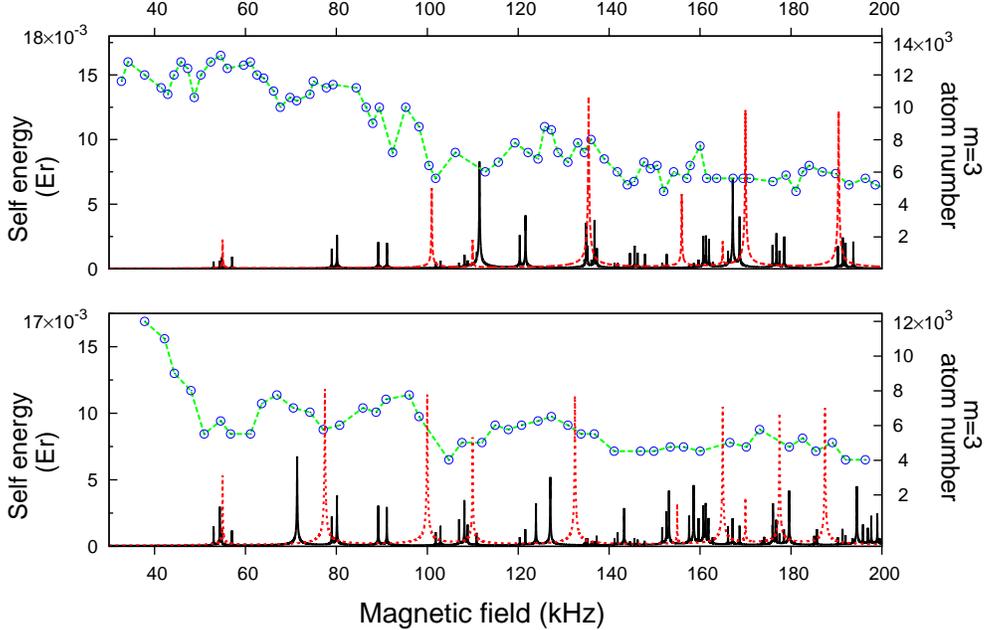}
\end{center} 
\caption{Comparison with the experiment: The self energy as a function of the Larmor frequency for 
the rectangular lattice of anisotropic sites - black lines. The assumed here lattice potential is 
separable and in every spatial direction has a form of a $`\sin^2$' function. Quadratic terms
in the expantion of the potential correspond to the following harmonic frequencies: 
$\omega_x/ 2\pi =  160$kHz, $\omega_y/2\pi  =  60$kHz
and $\omega_z/2\pi  = 90$kHz.  Two orthogonal orientations of the magnetic field
in the horizonatal plane, similarly as in the experiment, are considered: $y$ direction -- 
upper panel, $x$ direction -- lower panel. Occupation of the initial
$m=3$ state as measured experimentaly in \cite{Tolra_rezonanse} is shown for comaprison -- circles.
The occupation should exhibit a local minimum at every resonance. 
The red dashed line correspods to the self energy obtained from the harmonic approximation,
$\omega_x/ 2\pi =  170$kHz, $\omega_y/2\pi  =  55$kHz and $\omega_z/2\pi  = 100$kHz, 
without any two-body interactions as used in \cite{Tolra_rezonanse}.    
Trap frequencies    }\label{Fig8}
\end{figure*}

Resonant demagnetization of initially polarized $^{52}Cr$ atoms in $|m_S=3\rangle$ state
in optical lattice at ultra low magnetic fields has been observed recently \cite{Tolra_rezonanse}. 
This process is possible because of the dipole-dipole interactions. 
Due to a geometrical configuration of laser beams the lattice has quite
complicated geometry and results presented in the previous sections
do not directly  apply to the experiment. 

In this section we present 
quite realistic calculations for parameters of the experiment \cite{Tolra_rezonanse}.
The lattice barrier in the horizontal plane is sufficiently high, $V_0=25 E_R$, therefore
the experimental system is in the Mott insulating state with two particle per
site. We neglect tunneling and possible lattice inhomogeneities so we do not consider
situations with three or more atoms per site.

In the horizontal plane the main axes of a single site
are tilted with respect to the lines connecting the lattice minima. 
The single site is anisotropic. In the harmonic approximation frequencies of
a site potential  are $\omega_x/ 2\pi =  170(10)$kHz, $\omega_y/2\pi  =  55(5)$kHz
and $\omega_z/2\pi  = 100(10)$kHz.  The realistic potential is non separable
and cannot be approximated
by the sum of `sine square' functions as studied here. In particular the expansion
of the potential around the minimum gives, in addition to quadratic, also
the third order terms. They are not present in the expansion of the `sin square' potential. 
Exact treatment requires construction of realistic Wannier functions. 
Such an approach goes beyond the scope of this paper. Instead we will fit theoretical results for 
the 3D lattice potential of the form:
\begin{equation}
\label{V_aniz}
V(x,y,z)=V_0 \big[ \sin^2(k_x x)+\sin^2(k_y y) \big] +V_z\sin^2(k_z z),
\end{equation}   
to the experimental data. 
In Eq.(\ref{V_aniz})  $V_0 = 25 E_R$ and $k_z=1$. The remaining parameters $k_x$, $k_y$, and 
$V_z$ are choosen to get the harmonic frequencies $\omega_x=2k_x\sqrt{V_0}$,
$\omega_y=2k_y\sqrt{V_0}$, $\omega_z=2\sqrt{V_z}$, in an agreement with the 
experimental arrangement.

The range of magnetic fields studied experimentally corresponds to the Larmor frequency 
$\hbar \omega_L = g \mu_B B < \hbar 2\pi \times 200$kHz. In this region there is a number of excited
orbital states which are accessible via dipole-dipole interactions. The selection rules following
from the particular form of the dipole-dipole interactions give limitations
on the states. They can be labeled by a number of excitation quanta 
of vibrational motion of the doublon $(N_x,N_y,N_z)$. 
For the magnetic field directed along the z-axis 
the selection rules are the following: i) If $\Delta {\cal M}=-1$ then
the quantum numbers $N_x$ and $N_z$  are odd but $N_y$ is even 
or $N_y$ and $N_z$ are odd while $N_x$ is even.
ii) For processes with $\Delta {\cal M}=-2$, the final states must be characterized 
by even values of the sum $N_x+N_y$ and even values of $N_z$. 
For the range of magnetic field studied experimentally doublon states of energies up to
$E_{fin}=(N_x\omega_x+N_y\omega_y+N_z\omega_z)E_R < \hbar (2 \pi \times 400)$kHz can be populated.
This condition combined with the selection rules gives 10 resonantly coupled states
for orientation of the external magnetic field parallel to the x-axis and 8 states
for the field orientation along the y-axis. Detailed specification of the final states is given in
\cite{Tolra_rezonanse}.  

In the harmonic trap only the relative orbital motion of two interacting
atoms can be excited therefore the final state is unique. This is no longer true
in realistic anharmonic potentials. The quantum numbers 
$(N_x,N_y,N_z)$ define an energy band. The excitation energy is shared by two
atoms, i.e. $N_i=n_i+n^{\prime}_i$, ($i=x,y,z$), where $n_i$ 
and $n^{\prime}_i$ are excitation quanta of individual atoms. 
Within the given energy band there are many  two particle states (obeying the bosonic symmetry)
which can be coupled by the dipole-dipole interactions 
to the initial state. To find the positions of dipolar resonances as well as their 'strength'
we diagonalize the full Hamiltonian of the two particle system in the basis
of two-particle states which includes the initial state as well as all the states
belonging to all the energy bands within interesting energy range. We include all
inter- and intra-band contact interactions between particles in the basis states. 
The intra-band interactions are not negligible as the smallest band spacing
$\omega_y=2\pi \times 55$kHz is not much larger than the characteristic contact interactions
$\sim 2\pi \times 10$kHz. We include
$|2\rangle|2\rangle$ as well as $1/\sqrt{2}(|3\rangle|1\rangle+|1\rangle|3\rangle$
spin states for excited  system. This way we take into account a possible transfer of atoms
from the $|2\rangle|2\rangle$ to $|3\rangle|1\rangle$ spin space due to the
spin changing contact interactions.

The results for two different orientations of the magnetic field are presented
in Fig.\ref{Fig8}. To compare them with experimental
data of \cite{Tolra_rezonanse} we plot the self energy as a value of magnetic field (Larmor frequency).
The self energy displays a sharp maximum (black lines) at every resonant value of the magnetic field. 
The points correspond to the experimental data as copied from the paper \cite{Tolra_rezonanse}
(Fig.2) and indicate the number of atoms left in the initial $m=3$ state.
This quantity should be small at the resonance, i.e. minima in the experimental curve
ought to correspond to maxima of the self energy. For comparison we show the self energy in the harmonic
approximation with all contact interaction set to zero as shown in Fig.2 in \cite{Tolra_rezonanse}
-- dashed (red) line.

The first observation is that the very fine structures predicted theoretically are significantly
narrower than resonant features observed. They are simply not resolved in the experiment.
The reasons for this were discussed in the experimental 
paper and are attributed mainly to the increasing band-width for
higher energy lattices bands but also to the magnetic field and
lattice depth fluctuations which are broadening the resonances, at the
10 kHz level. The extremely precise control of magnetic field, probably
beyond the level of experimental reach is required.

The second observation is that the structure of resonances is very reach. This fact 
is obvious on the basis of results presented here:  
degeneracy of  every energy shell  is removed because 
of anharmonicity and anisotropy and  all the states become accessible
 via the dipole-dipole interactions.

Finally because of anharmonicity and the contact interactions the positions
of the resonances are shifted towards lower magnetic fields. Note that both above mentioned effects
act (mainly) in the same direction. 
The energy shift can be as large as $2\pi \times 10$kHz for the states excited to the high Wannier bands.
  
The anharmonicity and anisotropy of the lattice does not affect significantly
the widths of the resonances. Their main effect is to split the every resonant peak into a reach fine structure.
The width of an individual fine resonant feature is smaller than the width
predicted in the harmonic approximation, but its magnitude remains at the same level, i.e. $\sim 10\mu$G.
It is probably not possible, to resolve these fine structures experimentally. 
On the other hand, splitting of the singe resonant peak results in an overall broadening
of the entire structure composed of several fine features. The broadening due to this
splitting is given by the contact interactions strength and corresponds to the  magnetic
fields $\sim 1-10$mG. A precise control of the magnetic field at this level is quite realistic.
The widths of observed in  \cite{Tolra_rezonanse} resonant features ($\sim 10$kHz) correspond
to the above estimation and, according to our analysis, can be attributed to the effects
of anharmonicity and anisotropy of the binding potential.    

The resonant character of the process is visible in a region of low Larmor frequencies only.
At the first glance there are only few distinct experimental resonances in the region 
of small Larmor frequencies. By choosing 
the following values of the site frequencies: $\omega_x/2\pi = 160$kHz, $\omega_y/2\pi = 60$kHz,
$\omega_z/2\pi = 110$kHz, we can  reproduce reasonably well
the positions of the three first minima for the case with the
magnetic field directed along the x-axis. The similar agreement we get for the orientation 
of the magnetic field along the y-axis.
In the region of large $\omega_L$ where highly excited states become occupied,
resonances are relatively close to each other
and their number increases. Different resonant structure start to overlap. Therefore
at higher magnetic fields, due to the fine splitting of neighboring
resonances  the resonant character of the process 
should be lost. This is clearly seen in the experimental data for large $\omega_L$.

The fitted frequencies have values within the given range of experimental uncertainties
however they are different than the  harmonic frequencies used in the experimental paper to get
the red dashed lines in Fig.\ref{Fig8}. 
The positions of experimentally resolved resonances can be reproduced theoretically 
regardless whether anharmonicity and anisotropy of the lattice potential are taken into account
or not. Small uncertaintities of the lattice parameters, as well as their fluctuations,
give  sufficient freedom 
to  allow for fitting positions of the resonances using both a simple harmonic theory without any two-body interactions
as well as quite advanced approach accounting for all interactions and 
anharmonicity and anisotropy of the lattice sites.  However, the overall widths of the entire fine structures cannot
be reproduced in the harmonic approximation.

\section{Final Remarks}
\label{conclusions}

We have shown that the dipole-dipole interactions are very effective
probe of the two-body states of Chromium atoms in the external
trapping potential. They are very sensitive to anharmonicity and
symmetry of the single site. The anharmonicity of the single
trap becomes very important in
particular when higher Wannier states with non vanishing angular
momentum are concerned. The contact interactions between the
degenerate states in the site lift the degeneracy and
introduce a fine structure of the energy spectrum of the
system. Spectroscopy of this fine structure can
be done with the help of the dipolar resonances. Observation of spin
dynamics driven by the contact interactions is a precise method of
determination of contact interaction for both Fermi and Bose
atoms \cite{Bloch, Sengstock}. We show that spin dynamics due to
dipolar interactions at the resonant
magnetic field is yet another perfect tool for precise
determination of the scattering lengths of involved atoms. This
can be very important for measurement of the various scattering
length for dipolar elements such as Dysprosium or Erbium \cite{Lev,Ferlaino}.
Generalization of the approach presented here to
more atoms per site or Fermi species is straightforward.

\acknowledgments 
We are grateful to B. Laburthe-Tolra, P. Pedri, and Jan Mostowski
for helpful discussions. The work was supported by the 
(Polish) National Science 
Center grants No. DEC-2011/01/D/ST2/02019 (JP, TS), DEC-2011/01/B/ST2/05125 (MB),
and DEC-2012/04/A/ST2/00090 (MG),
Spanish MINCIN project TOQATA (FIS2008-00784),
EU IP AQUTE  and ERC Advanced
Grant QUAGATUA. T.S. acknowledges support from the Foundation 
for Polish Science (KOLUMB Programme; KOL/7/2012) and 
hospitality from ICFO.

\end{document}